\documentclass[traditabstract,letter]{aa}
\usepackage{graphicx}
\usepackage{natbib}
\usepackage{color}
\usepackage{txfonts}
\def\gr{$\gamma$-ray}

\begin{document}
\title{Gamma ray emission from magnetized relativistic GRB outflows}
\author{A.~Neronov and V.~Savchenko}
%\titlerunning{Gamma ray emission from magnetized relativistic GRB outflows}
%\authorrunning{A.~Neronov and V.~Savchenko}

\institute{ISDC Data Center for Astrophysics, Chemin d'Ecogia 16, 1290 Versoix,
Switzerland and Geneva Observatory, 51 ch. des Maillettes, CH-1290 Sauverny,
Switzerland}

\abstract {We argue that small pitch angle synchrotron emission provides an
  important dissipation mechanism which has to be taken into account in the
  models of formation of relativistic magnetized \gr\ burst (GRB) outflows from
  the newborn black holes and/or magnetars. We show that if the GRB outflow is
  proton loaded, spectral energy distribution of this emission is expected to
  sharply peak in 0.1-1~MeV energy band. If the small pitch angle synchrotron
  emission efficiently cools relativistic particles of the outflow, its spectrum
  below the peak energy is a powerlaw with spectral index $\alpha\simeq -1$,
  close to the typical spectral index of the time-resolved GRB
  spectra. Otherwise, the low energy spectral index can be as hard as
  $\alpha\simeq 0$, as observed at the beginning of the GRB pulses. We make a
  conjecture that small pitch angle synchrotron emission from proton-loaded
  magnetized GRB outflow could significantly contribute to the Band component of
  the prompt emission of GRBs while electromagnetic cascade initiated by the
  protons could be responsible for the GeV component.}


\maketitle

%%%%%%%%%%%%%%%%%%%%%%%%%%%%%%%%%%%%%%%%%%%%%%%
\textit{Introduction.}
%%%%%%%%%%%%%%%%%%%%%%%%%%%%%%%%%%%%%%%%%%%%%%
Over the last 2 decades a large amount of information about time evolution of
the spectral characteristics of prompt emission of \gr\ bursts (GRB) has
been collected by {\it CGRO}/BATSE, {\it Swift}/BAT, {\it INTEGRAL}/ISGRI (see
\citep{mesaros06,piran04} for recent reviews) and, most recently, {\it
  Fermi}/GBM \citep{band09} telescopes. It is firmly established that the
majority of the time resolved GRB spectra could be well described by relatively
simple spectral models, such as the Band model \citep{band93}, or a cut-off or
broken powerlaw models determined by three main parameters: low and high energy
photon indices $\alpha$ and $\beta$ and peak (or break or cut-off) energy
$E_p$ \citep{piran04,mesaros06}.

The observed distribution of the low-energy power-law indices $\alpha$ (of
differential photon spectra $dN_\gamma/dE\sim E^{\alpha}$) of time-resolved GRB
spectra is sharply peaked at the value $\alpha\simeq -1$ in the BATSE GRB
\citep{band93,kaneko06} and {\it Swift}/BAT \citep{savchenko09} GRB samples.
Deviations from this mean value toward $\alpha>-1$ (down to $\alpha\simeq 0$)
are preferentially observed at the initial moments of on-set of GRB sub-pulses,
when the spectra could be as hard as $\alpha\gtrsim 0$
\citep{preece00,kaneko06,savchenko09}. Evolution toward $\alpha<-1$ is commonly
observed during decay of individual sub-pulses and/or at the end of the prompt
GRB emission phase. The characteristic value $\alpha\simeq -1$ is difficult to
reconcile with the conventionally assumed synchrotron and/or inverse Compton
mechanisms of prompt GRB emission. Alternative models, which try to resolve this
problem, include ``jitter" radiation \citep{medvedev99,medvedev06} (see
\citet{kirk09} for discussion of potential problems of the jitter radiation
model) and ``quasi thermal comptonization" \citep{ghisellini99} models.

Spectra of at least some GRBs contain a separate GeV-band component
(e.g. \cite{grb080916c,bissaldi09}). The GeV emission has different (compared to
the Band component) temporal evolution \citep{grbgevradfire}. Absence of the high-energy cutoff in the
spectrum of this GeV component due to the pair-production imposes a lower
bound on the bulk Lorentz factors of the GRB outflows (about $\Gamma \gtrsim
1000$, e.g. \cite{baring97,grb080916c}). Different models of the origin of the
GeV emission were proposed, such as high-energy extention of the conventional
afterglow \citep{kumar09,Ghirlanda09}, afterglow emission from a fireball in the
radiative regime \citep{grbgevradfire}, synchrotron \citep{razzaque09} or
cascade \citep{dermer06,asano09} emission from ultra high-energy ($\sim
10^{20}~eV$) protons.

Several models of formation of relativistic GRB outflows from black holes or
neutron stars, formed in result of core collapse of massive stars, assume that
the outflow is highly magnetized
\citep{komissarov07,bucciantini07,komissarov09,bucciantini09}. In these models,
the energy of the outflow is initially carried by the Poynting flux of
electromagnetic energy, so that the ratio of electromagnetic to kinetic energy
of the outflow, $\sigma$, is comparable or much higher than unity. Numerical MHD
calculations show that the electromagnetic energy could be converted in a
dissipationless way to the kinetic energy of relativistic particle outflow. The
absence of dissipation is due to the fact that electromagnetic field forms a
force-free configuration of electromagnetic field, $\vec E+[\vec V\times \vec
B]=0$, where $\vec E,\vec B$ are electric and magnetic field and $\vec V$ is the
flow velocity.

In what follows we show that even if electromagnetic field is force-free, the
energy loss of relativistic particles in the outflow is non-zero, because
directions of velocities of individual particles are scattered around the
average direction of the flow within a certain angle $\Psi_0\sim \Gamma_0^{-1}$,
where $\Gamma_0$ is the Lorentz factor of the outflow. Scatter of particle
velocities leads to radiative dissipation of the outflow energy, with the main
dissipation mechanism being small pitch angle synchrotron emission.  We
demonstrate that significant fraction of the power of magnetized outflow could
be dissipated via this mechanism already in the outflow acceleration region.

We derive spectral characteristics of this emission and show that they are
consistent with those of the observed GRB prompt emission spectra.  The peak
energy of the spectrum is expected to be in the 0.1-1~MeV range for the outflows
with typical GRB parameters (bulk Lorentz factors, isotropic luminosities). The
low-energy spectral index is expected to evolve from a very hard value
$\alpha\simeq 0$ toward a "steady state" value $\alpha=-1$ which is expected to
persist over the period of stable activity of GRB central engine and/or up to
the moment when interactions of the high-energy protons with the
synchrotron-cooled lower energy protons start to significantly contribute to the
energy loss. Comparing the spectral characteristics of proton synchrotron
emission with the ones observed in the prompt emission \gr\ spectra of GRBs we
put forward a conjecture that emission from the pre-shock outflow dominates
during the prompt emission phase.  We discuss possible observational tests which
would allow to test this conjecture.

%%%%%%%%%%%%%%%%%%%%%%%%%%%%%%%%%%%%%%%%%%%
\textit{Generic properties of pre-shock relativistic GRB outflow.}
%%%%%%%%%%%%%%%%%%%%%%%%%%%%%%%%%%%%%%%%%%%
If the observed isotropic luminosity of GRB is $L_{\rm iso}$, the energy flux
carried by the GRB outflow ejected into a solid angle $\Omega$ is
$L=(\Omega/4\pi)L_{\rm iso}\simeq 8\times 10^{47}\left[L_{\rm
    iso}/10^{52}\mbox{erg/s}\right] \left[\Theta/1^\circ\right]^2\mbox{ erg/s}$,
where $\Theta$ is the opening angle of the outflow.  Energy density of the
outflow at a distance $D$ is $U=L/(\Omega D^2)$ \footnote{We use natural system
  of units $c=1$.}. It is convenient to parametrize the strength of
electromagnetic field in the outflow in terms of magnetization parameter
$\sigma$, $B^2/(8\pi)\sim\sigma U/(1+\sigma)$.  Using such a parametrization,
(electro)magnetic field strength at the distance $D$ can be expressed as
$B=\sqrt{ 2\sigma L_{\rm iso}/(1+\sigma)}/D\simeq
10^{13}\left[\sigma/(1+\sigma)\right]^{1/2}\left[L_{\rm iso}/10^{52}\mbox{
    erg/s}\right]^{1/2} \left[D/10^8\mbox{ cm}\right]^{-1}\mbox{ G}$.

Minimal possible angular scatter of velocities of the particles is within a cone
with opening angle $\Psi_0\simeq \kappa \Gamma_0^{-1}$ ($\kappa$ is a numerical
factor of the order of $1$). Non-zero angular scatter of particle velocities
around the bulk velocity $\vec V$ leads to the appearance of non-zero Lorentz
force of the order of $F\sim eB_\bot$, where $e$ is particle charge and
$B_\bot\sim B\Psi_0=\kappa B\Gamma_0^{-1} \simeq
10^{9}\kappa\left[\sigma/(1+\sigma)\right]^{1/2} \left[L_{\rm
    iso}/10^{52}\mbox{erg/s}\right]^{1/2} \left[D/10^8\mbox{cm}\right]^{-1}
\left[\Gamma_0/10^{4}\right]^{-1}\mbox{G.}$

Particle motions due to this uncompensated Lorentz force lead to radiation. To
calculate the properties of this radiation, it is convenient to go to the
reference frame moving with velocity ${\cal V}=\left[\vec E\times \vec
  B\right]/\left| B\right|^2$. In this reference frame the electric component of
electromagnetic field is zero, the mean outflow velocity $\vec V'$ is aligned
with the magnetic field $\vec B'$ and the radiation is synchrotron radiation of
particles moving at small pitch angles in magnetic field\footnote{In a special
  case $\vec V\bot\vec B$ the reference frame moving with velocity ${\cal V}$ is
  comoving with the outflow and the radiation is cyclotron radiation.}. Unless
$\vec V$ is orthogonal to $\vec B$, the velocity of the new frame ${\cal V}\ll
c$. Taking this into account, we will not distinguish the quantities (such as
photon energies) in different reference frames in the order-of-magnitude
estimates presented below.
 
Properties of synchrotron emission from particles moving at small pitch angles  with respect to the magnetic field are significantly different from those of synchrotron emission from an isotropic particle distribution \citep{epstein73,lloyd00}. The typical energy of the radiated photons, $\epsilon_s\sim eB_\bot \Gamma_0^2/m_p$  
\begin{equation} 
\label{psynch}
\epsilon_s \simeq 6\times 10^{5}\kappa\left[\frac{\sigma}{1+\sigma}\right]^{1/2}\left[\frac{L_{\rm iso}}{10^{52}\mbox{ erg/s}}\right]^{1/2}\left[\frac{D}{10^8\mbox{ cm}}\right]^{-1}
\left[\frac{\Gamma_0}{10^4}\right]\mbox{ eV}
\end{equation} 
where $m_p$ is particle mass (proton mass in the above and in the following numerical estimates), scales proportionally to the particle energy, rather than to the square of it. The synchrotron energy loss rate 
$P_s= 2e^4B_\bot^2\Gamma_0^2/(3m_p^2)\sim \kappa^2B^2$  does not depend on particle energies $\Gamma_0m_p$.

In order to estimate the importance of small pitch angle synchrotron cooling as a dissipation mechanism, one has to compare the synchrotron cooling distance  $D_s=\Gamma_0m_p/P_s\simeq 10^7\kappa^{-2} \left[\sigma/(1+\sigma)\right]^{-1}\left[L_{\rm iso}/10^{52}\mbox{erg/s}\right]^{-1} \left[D/10^8\mbox{cm}\right]^{2}\left[\Gamma_0/10^{4}\right]\,\mbox{cm}$ with typical distance scales of formation and propagation of the GRB outflow. Equating $D_s= D$ one can find that synchrotron emission can efficiently remove energy from relativistic protons up to the distance 
\begin{equation}
\label{D_p}
{\cal D}_{s}\simeq 1.3\times 10^{9}\kappa^2\left[\frac{\sigma}{1+\sigma}\right]\left[\frac{L_{\rm iso}}{10^{52}\mbox{ ergs/s}}\right]\left[\frac{\Gamma_0}{10^{4}}\right]^{-1}\ \mbox{ cm}
\end{equation}
Typical distance scale of the GRB central engine $R_{\rm CE}\sim 10^6..10^7$~cm is the size of black hole formed in the collapse of massive star, or the radius of the magnetar. The region of acceleration of GRB outflow to its asymptotic Lorentz factor $\Gamma_0$ might have much larger size, e.g.  $D_{\Gamma}\sim \Gamma_0R_{\rm CE}$ in the ``fireball" model \citep{mesaros06}.{\bf A similar estimate of the size of the GRB outflow acceleration region is found in the ``magnetic acceleration'' type models \citep{komissarov07,komissarov09}.}   Both $R_{\rm CE}$ and $D_\Gamma$ might be $\lesssim {\cal D}_s$ so that small pitch angle synchrotron emission  can be efficient dissipation mechanism. 

{\bf In the fireball model it is commonly assumed that the bulk of the power of the GRB outflow is transfered from the region of the outflow formation up to the internal shock region by relativistic protons, while the fraction of the power carried by electrons is smaller by a factor $\sim m_e/m_p\simeq 10^{-3}$ \citep{mesaros06}. Power of the proton-loaded outflow is efficiently transfered to electrons only in the internal shock region, which is normally at large distances $D_{\rm shock}\sim \Gamma^2R_{\rm CE}\gg {\cal D}_s$. In the unshocked outflow, electrons suffer, similarly to protons, from the energy loss related to the small pitch angle synchrotron emission. However, the energy losses of electrons do not contribute significantly to the energy dissipation in the distance range $D_\Gamma<D<D_{\rm shock}$.}

Synchrotron photons could freely escape from the GRB outflow if they are produced at sufficiently large distances, where the GRB outflow is optically thin with respect to Compton scattering. The mean free path of photons with respect to the Compton scattering is $\lambda_{\rm C}\simeq \left[n\sigma_T(1-\cos(\Psi_0))\right]^{-1}\simeq2\times 10^9\kappa^{-2}\left[D/10^8\mbox{ cm}\right]^2\left[L_{\rm iso}/10^{52}\mbox{ erg/s}\right]^{-1} \left[\Gamma_0/10^4\right]^{3}\mbox{ cm}$,
where $\sigma_T$ is Thomson cross-section and $n=\left.L_{\rm iso}\right/(4\pi D^2 m_p\Gamma_0)$ is the electron density of the outflow (we assume that the number density of electrons is equal to the number density of protons). The outflow becomes optically thin at the distance at which  $\lambda_{\rm C}\simeq D$:
\begin{equation}
{\cal D}_{\rm C}\simeq 5\times 10^{6}\left[\frac{L_{\rm iso}}{10^{52}\mbox{ erg/s}}\right]\left[\frac{\Gamma_0}{10^4}\right]^{-3}\mbox{ cm}
\end{equation}
In GRBs with bulk Lorentz factors $\Gamma_0\ge 6\times 10^2$, ${\cal D}_{\rm C}<{\cal D}_s$ so that synchrotron photons freely escape if they are produced in the distance range ${\cal D}_{\rm C}<D<{\cal D}_s$. In the distance range $D<{\cal D}_{\rm C}$ the spectrum of the proton synchrotron radiation is modified by the small angle Compton scattering on (energetically sub-dominant) electrons present in the outflow. 

%%%%%%%%%%%%%%%%%%%%%%%%%%%%%%%%%%%%%%%%%%
\textit{Spectrum of emission from pre-shock proton loaded GRB outflow}
%%%%%%%%%%%%%%%%%%%%%%%%%%%%%%%%%%%%%%%%%%%%
 could be calculated by summation the spectra emitted from the outflow components moving at different angles $\theta$ with respect to the line of sight 
 \begin{equation}
 \label{integral}
 F_{\Gamma_0}(\nu)=\frac{1}{\Psi_0^2}\int  \theta d\theta \int_0^{\Psi_0}\Psi d\Psi F_{\Gamma_0,\theta}(\nu,\Psi)
 \end{equation}
 where $F_{\Gamma_0,\theta}(\nu,\Psi)$ is emission from particles spiraling at pitch angle $\Psi$ within a beam with an opening angle $\Psi_0= \kappa\Gamma_0^{-1}$ viewed at an angle $\theta$ with respect to the direction of the beam. Non-negligible contribution to the integral over $\theta$ is given only by $\theta$ in the range $0<\theta<(\Psi_0+\Gamma^{-1})$. Note that we have assumed, for simplicity, that the parameters of the outflow do not depend on the angle with respect to the axis of the outflow, i.e. the Lorentz factor $\Gamma_0$ and particle density $n$ are constant inside a cone with large enough opening angle $\Theta\gg \Gamma^{-1}$. 
 Generalization of the calculations presented below to the case of non-trivial angular profile of the GRB outflow is straightforward.
 
\begin{figure}
\includegraphics[width=\linewidth]{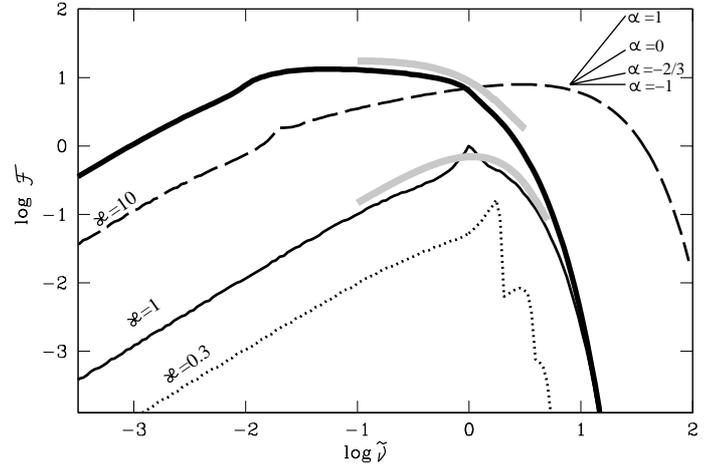}
\caption{Spectra of small pitch angle proton synchrotron emission from pre-shock GRB outflow for several values of $\kappa$. Thick solid line shows spectrum of emission from proton spectrum formed by cooling of the outflow protons to $\Gamma=0.01\Gamma_0$. For comparison we show Band model fits to the spectra of periods "b" (beginning) and "d+e" (end) of {\it Fermi} GRB 090902B \citep{bissaldi09} as lower and upper grew curves. Normalizations and peak energies of the Band model fits to GRB 090902B spectra are shifted to match the small pitch angle synchrotron model curves.}
\label{fig:spectra}
\end{figure}

$F_{\Gamma_0,\theta}(\nu,\Psi)$ in Eq. (\ref{integral}) could be expressed as \citep{ginzburg69,epstein73} 
 \begin{eqnarray}
 F_{\Gamma_0,\theta}(\nu,\Psi)&\simeq& \frac{16\pi e^2\nu_B^2\Gamma_0^4\Psi^2}{(1+\theta^2\Gamma_0^2+\Psi^2\Gamma_0^2)^3}\sum_{n=1}^\infty n^2\left[(J'_n(Z_n))^2\right.\nonumber\\&&\left.+\left\{\frac{1-\theta^2\Gamma_0^2+\Psi^2\Gamma_0^2}{2\theta\Psi\Gamma_0^2}J_n(Z_n)\right\}^2\right]
 \delta(\nu-\nu_n)\nonumber
 \end{eqnarray}
where $\nu_B=eB/(2\pi m)$ is the cyclotron frequency,
$\nu_n=2n\Gamma_0\nu_B/(1+\theta^2\Gamma_0^2+\Psi^2\Gamma_0^2)$,  $Z_n=2n\Gamma_0^2\theta\Psi/(1+\theta^2\Gamma_0^2+\Psi^2\Gamma_0^2)$ and $J_n(z)$ are Bessel functions. Substituting the above expression into Eq. (\ref{integral}) one finds
\begin{equation}
\label{F_final}
F_{\Gamma_0}(\nu)=\frac{4\pi e^2\nu_B }{\kappa^2\Gamma_0}{\cal F}(\tilde\nu,\kappa)
\end{equation}
where $\tilde\nu=\nu/(2\Gamma_0\nu_B)$ and ${\cal F}(\tilde \nu,\kappa)$ is a function which does not depend on the magnetic field strength and on the Lorentz factor. Numerically calculated functions ${\cal F}(\tilde\nu,\kappa)$ are shown for several values of $\kappa$ in  Fig. \ref{fig:spectra}. At low frequencies, $\tilde\nu\ll 1$, the function ${\cal F}_\gamma(\tilde\nu,\kappa)$ has the form of a hard spectral index power-law, 
${\cal F}_\gamma(\tilde\nu,\kappa)\sim \tilde\nu^{\alpha+1}$ with  $\alpha=0$.
The spectrum  becomes even harder close to the Doppler shifted cyclotron frequency $\tilde\nu=1$.  If $\kappa<1$ the function ${\cal F}(\tilde\nu,\kappa)$ has more-or-less strong higher harmonics of the cyclotron line which can be seen as multiple "bumps" in the spectrum at $\tilde\nu\ge 1$.

%%%%%%%%%%%%%%%%%%%%%%%%%%%%%%%%%%%%%%%%%%%%
\textit{High-energy emission from the "tail" of angular distribution.}
%%%%%%%%%%%%%%%%%%%%%%%%%%%%%%%%%%%%%%%%%%%%%
The spectrum marked $\kappa=1$ in Fig. \ref{fig:spectra} is calculated under the assumption that angular distribution of particle velocities, $N_p(\Psi)$, is cut-off at the angle $\Psi_0\simeq \Gamma_0^{-1}$. In general, $N_p(\Psi)$ could have large angle "tail" extending beyond $\Psi\simeq \Gamma_0^{-1}$. Spectrum of emission from protons moving at large pitch angles $\kappa\gg 1$ extends up to the frequency $\tilde\nu\simeq \kappa$, while the frequency below which the spectrum is characterized by $\alpha=0$ moves down to $\tilde\nu\simeq \kappa^{-2}$ (see Fig. \ref{fig:spectra}). In the range $\kappa^{-2}\tilde\nu<\kappa$ the spectrum is characterized by photon index $\alpha=-2/3$ typical for synchrotron radiation from isotropic particle distribution (Fig. \ref{fig:spectra}). If  the large pitch angle tail of $N_p(\Psi)$ has the form of a power-law, $N_p(\kappa\Gamma_0^{-1})\sim \kappa^{-s}$, the spectrum of emission above the characteristic (peak) frequency $\nu=2\Gamma_0\nu_B$ has the form of a power-law with photon index $\beta=s-1$. In principle, the tail of the angular distribution could extend up to $\Psi\sim 1$, so that the energy of synchrotron photons emitted by the large angle tail could extend up to $\tilde\nu\sim\Gamma_0$. If the spectrum of proton synchrotron emission from particles moving at the angle $\Psi\le \Gamma_0^{-1}$ peaks in the 100 keV -- 1 MeV energy band, the high-energy (large pitch angle) tail of the synchrotron spectrum could extend up to the 1 -- 10 GeV band if $\Gamma_0\sim 10^4$. Presence of the large angle tails of proton angular distribution could, therefore, be responsible for the appearance of high-energy extensions of the prompt GRB spectra like the one observed in GRB 080916C \citep{grb080916c}. Another obvious possible source of the GRB 080916C-like high-energy tails in the GRB spectra is presence of protons with energies higher than $\Gamma_0m_p$ in the outflow. 

%%%%%%%%%%%%%%%%%%%%%%%%%%%%
%%%%%%%%%%%%%%%%%%%%%%%%%%%%

%%%%%%%%%%%%%%%%%%%%%%%%%%%%%%%%%%%%%%%%%%%%
\textit{Emission from a distribution of protons formed by the small pitch angle synchrotron cooling.}
%%%%%%%%%%%%%%%%%%%%%%%%%%%%%%%%%%%%%%%%%%%%%
The total proton energy loss rate could be found via integration of $F_{\Gamma_0}(\nu)$:
\begin{equation}
P_{\rm synch}(\Gamma_0)=\int F_{\Gamma_0}(\nu) d\nu=\frac{8\pi e^2\nu_B^2}{\kappa^2}\int {\cal F}(\tilde\nu,\kappa)d\tilde\nu
\end{equation}
As expected,  $P_{\rm synch}$ does not depend on the Lorentz factor $\Gamma_0$. Energy-independent small pitch angle synchrotron cooling leads to generation of low-energy tail of particle spectrum below the initial proton energy $\Gamma_0m_p$. During the time when the central engine of the GRB outflow continuously ejects relativistic outflow with rate $Q(\Gamma)=Q_0\delta (\Gamma-\Gamma_0)$, the spectrum of the synchrotron-cooled protons $N_p(\Gamma)$ is a stationary solution of continuity equation $\left.\partial\right/\partial \Gamma\left\{P_{\rm synch}(\Gamma)N_p(\Gamma)\right\}=Q(\Gamma)$:
\begin{equation}
\label{b}
N_p(\Gamma)=\frac{Q_0}{P_{\rm synch}(\Gamma)}\sim \Gamma^{-b},\ \ \ b=0
\end{equation}
in the range of Lorentz factors $\Gamma\le \Gamma_0$.
Spectrum of small pitch angle synchrotron emission from such proton distribution is
\begin{equation}
\label{steady}
F(\nu)=\int N_p(\Gamma) F_\Gamma(\nu)d\Gamma\sim\nu^{\alpha+1}, \alpha=-(b+1)=-1
\end{equation}

The overall spectral evolution of the pre-shock outflow emission is expected to be as follows. As soon as synchrotron cooling of the outflow particles becomes efficient, the spectrum of emission from the pre-shock outflow softens from $\alpha\simeq 0$ (or $\alpha=-2/3$, depending on the initial value of $\kappa$) down to $\alpha\simeq -1$. The $\alpha\simeq -1$ spectrum persists over a period of stable activity of  the GRB central engine as long as synchrotron cooling remains the dominant energy loss channel.  This value of $\alpha$ is close to the measured characteristic low-energy spectral index of {\it CGRO}/BATSE and {\it Swift}/BAT GRBs \citep{kaneko06,savchenko09}. Prompt emission spectra of some GRBs start from very low values of $\alpha>-2/3$ \citep{preece00,kaneko06,savchenko09} and evolve toward $\alpha\simeq -1$ as expected if the very hard emission originates from the pre-shock GRB outflow. As an example, we show in Fig. \ref{fig:spectra} Band model fits to the spectra of the beginning and end of the prompt emission phase of a recent GeV-\gr -loud GRB 090902B, taken from \citep{bissaldi09}. Comparing the Band model fit of the initial rising phase of GRB 090902B with the spectrum of emission from monochromatic proton distribution one could conclude that GRB 090902B outflow is initially almost monoenergetic. Eq. (\ref{psynch}) enables to derive initial $\Gamma_0\sim 10^4$, i.e. proton energy $\Gamma_0m_p\sim 10^{13}$~eV, given the isotropic luminosity $E_{\rm iso}\sim 10^{53}$~erg/s and peak energy $E_{\rm peak}\sim \epsilon_s\sim 10^6$~eV for this particular GRB.

Small pitch angle synchrotron emission leads not only to cooling but also to scattering of proton beam into a wider angle $\Psi\sim \Gamma^{-1}\gg \Gamma_0^{-1}$. Widening of the opening angle of particle distribution could lead to the "switch on" of interactions between particles of the outflow. The mean free path of the highest energy protons with respect to the $pp$ interactions on low energy protons $\lambda_{pp}=\left\{\sigma_{pp}n(\Gamma) (1-\cos(\Gamma^{-1}))\right\}^{-1}\simeq 10^{11}\left[L_{\rm iso}/10^{52}\mbox{ erg/s}\right]\left[D/10^8\mbox{ cm}\right]\left[\Gamma_0/10^4\right]\left[\Gamma/10^4\right]$~cm where $\sigma_{pp}\simeq 10^{-26}$~cm$^2$ is the cross section of $pp$ interactions and $n(\Gamma)\sim \Gamma^{1-b}$ is the density of particles with gamma factor $\Gamma$. $\lambda_{pp}$ becomes comparable or shorter than ${\cal D}_s$ as soon as synchrotron cooling leads to the appearance of protons with energies $\Gamma\lesssim 10^{-2}\Gamma_0$ if $\Gamma_0\sim 10^4$. 

{\bf Development of proton initiated cascade leads to injection of a secondary electrons/positrons (as $\gamma$-rays and well as neutrinos) which could largely outnumber the primary electrons present in the unshocked GRB outflow. Contrary to the primary electrons, synchrotron emission from these cascade electrons could provide a significant energy dissipation mechanism of the outflow. Although typical initial Lorentz factors of the secondary electrons are $\gamma_e\sim (\Gamma m_p)/m_e\gg \Gamma_0$, electrons are injected at the typical pitch angles $\psi_e\sim \Gamma^{-1}\gg \gamma_e^{-1}$, so that the synchrotron emission from the cascade electrons is not emitted in the small pitch angle regime. } 

\gr s with energies above the pair production threshold 
$E_{\gamma\gamma}=\sqrt{2m_e^2c^4/(1-\cos(1/\Gamma))}\simeq 10^{10}\left[\Gamma/10^4\right]\mbox{ eV}$ will produce pairs in interactions among themselves and initiate development of electromagnetic cascade. Highest energy \gr s could interact also with protons via pion and pair production channels. The energy loss of protons due to the $pp$ (and/or $p\gamma$) interactions $P_{pp}(\Gamma_0)\sim f\Gamma_0/\lambda_{pp}$, where $f\sim 0.1-0.2$ is the typical inelasticity of $pp$ collisions. As soon as cooling of the highest energy protons via $pp$ and $p\gamma$ interactions becomes more efficient than the synchrotron cooling, the spectrum of protons is expected to soften to $b>0$ (see Eq. (\ref{b})). This should lead to the softening of the small pitch angle synchrotron emission spectrum to $\alpha<-1$ and decrease of its contribution to the overall GRB spectrum, compared to the contribution from the cascade emission component. The spectrum of emission from the proton-initiated cascade extends up to the energy of the threshold of the pair production, $E_{\gamma\gamma}$. In the particular example GRB 090902B, the electromagnetic cascade component is then clearly identified as the soft emission component extending up to the multi-GeV energies, which becomes dominant at the end of the prompt emission phase and persists over $\sim 10^3$~s after the end of the prompt emission \citep{bissaldi09}.

Bursts which start with very hard spectra would be most interesting candidates
for testing the hypothesis of small pitch angle proton synchrotron emission
because they might possess the cyclotron line at the frequency $\nu\sim
\Gamma_0\nu_B$. Search for such a line at the energy close to the peak energy of
very hard GRB spectra could provide a test for the proposed model. The cyclotron line feature could, however, be "washed out" of the spectrum by
non-negligible spread of particle energies in the outflow and by small angle
Compton scattering if most of the small pitch angle synchrotron emission is
generated at the distances $D\lesssim {\cal D}_{\rm C}$. In this case the
spectrum of emission from the pre-shock outflow would be indistinguishable from
the generic Band model spectra. A complementary way of testing of the small
pitch angle proton synchrotron model is the search for the appearance of
neutrino signal from $pp$ interactions \citep{paczynski94} at the moment of
softening of the spectrum of prompt emission to $\alpha\le 1$ in GeV \gr -loud
GRBs. The neutrino signal is expected to be sharply peaked at energies
$E_\nu\sim f\Gamma_0m_p\simeq 1\left[\Gamma_0/10^4\right]$~TeV and its flux is
expected to be comparable to the luminosity of the cascade (GeV) component of
the GRB spectrum. The peak energy of neutrino signal can be predicted if the
bulk Lorentz factor $\Gamma_0$ is estimated from the measurement of the peak
energy of the proton synchrotron component using Eq. (\ref{psynch}). Search for
the neutrino counterparts of the GeV \gr\ loud GRBs becomes possible via
cross-correlation of the signal of {\it Fermi}/LAT detected GRBs with the TeV
neutrino signal in km$^3$ scale neutrino telescope IceCube which will be
completed in the nearest future \citep{icecube}.

To summarize, we have explored the possibility that small pitch angle proton synchrotron emission from the magnetized GRB outflow gives significant contribution to the GRB spectrum. This emission provides an important dissipation mechanism in the region of acceleration of GRB outflows with high magnetization parameter $\sigma$. We have shown that steady-state spectrum of this emission is expected to have photon index $\alpha=-1$, close to the characteristic photon index of the time-resolved GRB spectra.  Small pitch angle proton synchrotron emission component could also explain  extremely hard spectra $\alpha\gtrsim 0$ observed at the beginning of some GRBs. The possibility that small pitch angle proton synchrotron emission from the region of acceleration of GRB outflow could be identified in the observed GRB spectra implies that the models of formation of magnetized relativistic outflow by newly born stellar mass black holes or magnetars could be observationally tested. The hypothesis of the presence of small pitch angle synchrotron emission component in the GRB spectra could be verified via search of the cyclotron line features in spectra of (some of the) hardest GRBs and/or via search of prompt TeV neutrino emission from GeV \gr -loud GRBs.  

\textit{Acknowledgment}. We would like to thank A.Taylor for discussions of the
subject. The work of AN is supported by the Swiss National Science Foundation
grant PP00P2\_123426.

\end{document}